\documentclass[
 amsmath,amssymb,
 aps,
]{revtex4-2}
\usepackage{float}
\restylefloat{figure}
\usepackage{graphicx}
\usepackage{dcolumn}
\usepackage{bm}
\usepackage{physics}
\usepackage{amsmath}
\usepackage{breqn}
\usepackage{subcaption}
\DeclareUnicodeCharacter{2212}{-}

\makeatletter
\let\cat@comma@active\@empty
\makeatother

\begin{document}

\title{\textbf{Measurement of gravitational acceleration in a single laser operated atomic fountain}}

\author{Kavish Bhardwaj\textsuperscript{1}}
\author{S. Singh\textsuperscript{1}}%
\author{S. P. Ram\textsuperscript{1}}
\author{B. Jain\textsuperscript{1}}
\author{Vijay Kumar\textsuperscript{1,3}}
\author{Ayukt Pathak\textsuperscript{2}}
\author{Shradha Tiwari\textsuperscript{2}}
\author{\\V. B. Tiwari\textsuperscript{1,3}}
\author{S. R. Mishra\textsuperscript{1,3}}
 \email{srm@rrcat.gov.in}
\affiliation{
 \textsuperscript{1} Laser Physics Applications Division, Raja Ramanna Centre for Advanced Technology, Indore - 452013, India.\\
 \textsuperscript{2} Laser Controls and Instrumentation Division, Raja Ramanna Centre for Advanced Technology, Indore - 452013, India.\\
 \textsuperscript{3} Homi Bhabha National Institute, Training School Complex, Anushakti Nagar, Mumbai - 400094, India.
}

\maketitle
\noindent

\section*{Abstract} We present measurements on Earth's gravitational acceleration (g) using an in-house developed cold atom gravimeter (CAG) in an atomic fountain geometry. In the setup, the laser cooled $^{87}Rb$ atoms are launched vertically up in the fountain geometry and Doppler sensitive two-photon Raman pulse atom interferometry is applied to detect the gravitational acceleration experienced by the atoms. Using our gravimeter setup, we have measured the local value of ‘g’ in our laboratory with sensitivity of 621 $\mu$Gal for integration time of 1350 s. 

\section{Introduction} 
For nearly three decades, the research and developments in field of cooling and trapping of neutral atoms have witnessed its applications ranging from fundamental science to advanced technologies \cite{inguscio}. Using laser cooled atoms, various fundamental physics problems, such as Bose Einstein Condensation \cite{BEC}, degenerate Fermi-gases \cite{Fermi}, equivalence principal \cite{Kasevich1999}, measurement of fine structure \cite{FineStructure}, measurement of fundamental constants \cite{rosi2014precision}, gravitational waves \cite{Tino}, etc, have been studied in more comprehensive way. The laser cooled atoms are now a days considered very promising candidates for developing state-of-art devices for next generation quantum technologies. These include various quantum sensors and metrology devices such as atomic clocks \cite{atomic}, atomic accelerometers \cite{inertial, car-gravimeter, PetersDropping, inertialsensor}, atom gyroscopes \cite{gyro, gyro2}, atomic sensors of electromagnetic fields \cite{microwave}, etc. Besides that, a considerable research and development work is also going on to make use of a single trapped atom as a quantum qubit for quantum information processing \cite{computing}. Due to involvement of quantum physics in working principle of devices, these cold atoms based quantum devices have shown inherent superiority as compared to existing state-of-art classical devices \cite{PRLgravityclassical,comparison}. 

After the first proposal and demonstration by Kasevich and Chu in 1991 \cite{KasevichSRT, kasevich1992measurement}, the cold atom interferometry based technique has become a very popular tool for precision measurement of inertial parameters such as acceleration \cite{Peters, rosi, Zhou, g_measurement, Gravimeter_dynamic_measurement, gravimetryreview, Fountainmicrowave} and rotation \cite{rotation}. A cold atom gravimeter (CAG), an instrument developed using cold atoms for precision measurement of earth's gravitational acceleration (g), can be more accurate than the conventional classical gravimeters. As compared to light interferometry based gravimeters, the CAG does not have any mechanical abrasion for the test mass used in measurements, which helps in increasing the repetition rate of measurements and maintaining device without frequent calibrations. Because of variations in underneath mass-density in the earth's body, the earth’s gravitational acceleration (g) varies from place to place on the earth's surface. The value of g varies roughly from 9.77$m/s^{2}$ to 9.84$m/s^{2}$ over the whole earth surface. It may also vary temporally during the underground seismic activities. The g measurements performed on the earth’s surface, using an accurate gravimeter, has various applications such as underground mineral explorations \cite{wu2019gravity}, monitoring seismic activities \cite{wu2014investigation}, detection of underground tunnels and structures, study of geophysics \cite{niebauer2007gravimetric}, etc. The data of accurate measurement of g, with its spatial and temporal variations,  is useful information for these applications. The accurate gravimeters have also applications in space technology such as inertial navigation and space missions for micro-gravity experiments \cite{yu2006development,tino2013precision}. The gravimeters based on cold atoms have shown high precision and accuracy due to involvement of fundamental principles of quantum mechanics in the measurements. The high accuracy of CAG originates from high precision measurement of a two-photon Raman transition in an atom. A CAG is an absolute gravimeter with high sensitivity, accuracy and long-term stability \cite{kasevich1992measurement,peters2001high}. 

In this work, we report the measurement of earth's gravitational acceleration (g) using an in-house developed cold atom gravimeter. This gravimeter is based on Doppler sensitive two-photon Raman excitation in cold $^{87}Rb$ atoms along with atom interferometry, while atoms are launched in vertically upward direction against gravity in an atomic fountain geometry. The atomic fountain reported here is similar to our earlier reported fountain based on a single laser system \cite{singh2021single}. The whole gravimeter setup operates with a single laser system to perform cooling of atoms, their launching in fountain, state-preparation and atom interferometric Raman excitations. Though the absorption probe beam for the state detection after interferometry was derived from another low power laser, to keep flexibility of locking it to different transitions, it was possible for us to obtain it from the same master laser. The measurement of g was performed when atoms were moving against gravity in the atomic fountain. From the measurements, we obtain the local value of ‘g’ in laboratory to be 9.786173 $m/s^{2}$ at the resolution of 270 $\mu$Gal ($\sim$ 0.000003 $m/s^{2}$), estimated from the frequency chirp rate applied to a Raman beam to attain a maximum in the interferometric fringes. In order to assess the performance of the system, the Allan deviation in the measurement of g was estimated to be 621 $\mu$Gal for integration time of 1350 s. The CAG setup was operated without implementing the seismic vibration corrections. 

\section{The principle of Raman pulse atom interferometer}

\begin{figure}[h]
    \centering
    \includegraphics[scale=0.35]{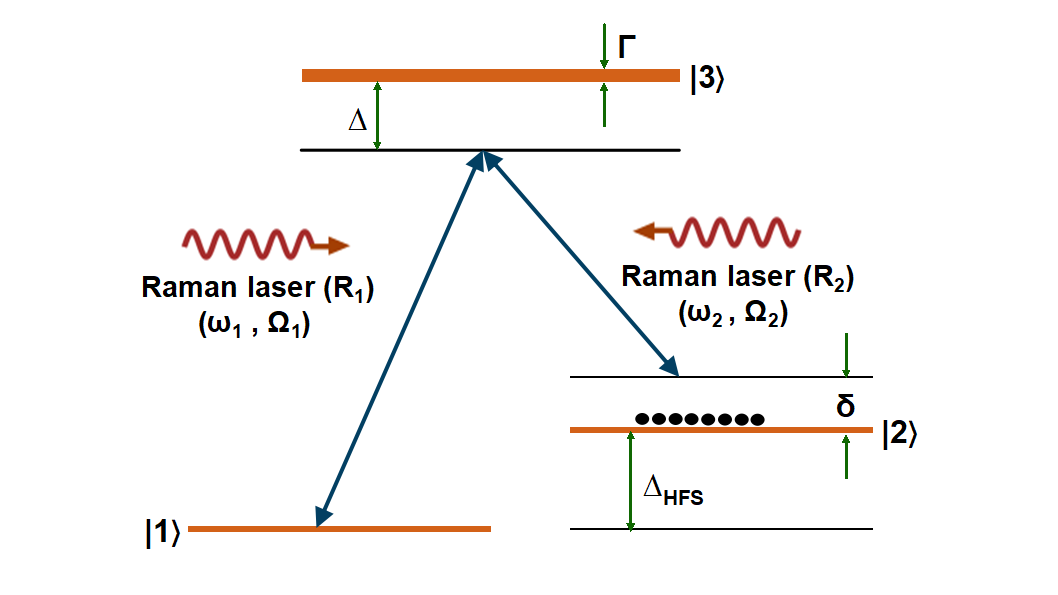}
    \caption{ The schematic of interaction of Raman laser beams with three level atom. The Raman beams ($R_1$ and $R_2$) have frequencies $\omega_{1}$ and $\omega_{2}$, and Rabi coupling strengths $\Omega_{1}$ and $\Omega_{2}$. The states $|1\rangle$ and $|2\rangle$ are coherently coupled by the Raman beams via an intermediate state $|3\rangle$ which is frequency detuned by $\Delta$. The value of $\Delta$ is much larger than the linewidth ($\Gamma$) of the excited state $|3\rangle$. The detuning $\delta$ is two-photon detuning of Raman beams, and $\Delta_{HFS}$ is the energy gap between two states $|1\rangle$ and $|2\rangle$.}
    \label{SRT}
\end{figure}

The basic principle, exploited in the CAG, is velocity sensitive two-photon Raman excitation in an atom along with the internal state atom interferometry. The details of this two-photon Raman transition phenomenon and its use for atom interferometry can be found in several earlier works \cite{theoreticalSRT, g_measurement, Zhou, Tinsley2019}. In this atom interferometry based g-measurement, an atom moving under the influence of gravitational acceleration g is probed using velocity sensitive two-photon Raman excitation between two internal states $|1\rangle$ and $|2\rangle$ via an intermediate state $|3\rangle$, as shown in Fig. \ref{SRT}. Because of changing speed of atom due to gravitational acceleration (g), the change in velocity sensitive (via Doppler effect) Raman transition probability is observed, which is used for estimation of g. 

In the atom interferometer, two Raman beams ($R_1$ \& $R_2$) at frequencies $\omega_{1}$ and $\omega_{2}$, with coupling strengths $\Omega_{1}$ and $\Omega_{2}$, are used to manipulate the populations in levels $|1\rangle$ and $|2\rangle$ (which are ground hyperfine states $|F = 1\rangle$ and $|F = 2\rangle$ in case of $^{87}Rb$ atom). When detuning ($\Delta$) is kept large enough to avoid any one-photon transition, this three-level system can be approximated to a two-level system with effective Rabi frequency for coupling of levels $|1\rangle$  and $|2\rangle$ as $\Omega_{eff}$=$\Omega_{1}$$\Omega_{2}$/$(2\Delta)$. If we consider multiple excited states instead of a single intermediate state $|3\rangle$, then the effective Rabi frequency ($\Omega_{eff}$) for Raman transition between two $|1\rangle$ and $|2\rangle$ is calculated by summing over all allowed intermediate excited levels (i) as \cite{kasevich1992},
\begin{equation}
    \Omega_{eff} = \sum_{i} {\frac{{\Omega}^*_{1i}\Omega_{i2}}{2 \Delta_{i}}}
    \label{SRT_Multiple}
  \end{equation}  
where the index i ranges over all intermediate allowed excited levels, $\Delta_{i}$ is detuning of Raman beams from level $|i\rangle$,  and $\Omega_{1i}$ and $\Omega_{2i}$ are single photon Rabi coupling strengths of each Raman beam for transition to level $|i\rangle$. 

The population amplitudes $C_{1}$(t) and $C_{2}$(t)) of levels $ |1\rangle$ and $|2\rangle$, at any time `t' during the interaction of Raman beams with atom, are given as \cite{Tinsley2019},

\begin{dmath}
    \label{eq:ground_pop1}
   C_{1}(t_0 + t) = \left[ \left\{\cos\left(\frac{\Omega_R t}{2}\right)-i\left(\frac{\delta-\delta_{AC}}{\Omega_R}\right)\sin\left(\frac{\Omega_R t}{2}\right)\right\}C_1(t_0) + \nonumber\\
   e^{i(\delta t_0+\phi_R)}\left\{-i\frac{\Omega_{eff}}{\Omega_R} \sin\left(\frac{\Omega_R t}{2}\right)\right\}C_2(t_0) \right]e^{-i(\Omega_1^{AC}+\Omega_2^{AC}-\delta)t/2}
\end{dmath}
and
\begin{dmath}
    \label{eq:ground_pop2}
C_{2}(t_0 + t) = \left[e^{-i(\delta t_0+\phi_R)}\left\{-i\frac{\Omega_{eff}}{\Omega_R} \sin\left(\frac{\Omega_R t}{2}\right)\right\}C_1(t_0)+ \nonumber\\    
   \left\{\cos\left(\frac{\Omega_R t}{2} \right)+i\left(\frac{\delta-\delta_{AC}}{\Omega_R}\right)\sin\left(\frac{\Omega_R t}{2}\right)\right\}C_2(t_0)\right]  e^{-i(\Omega_1^{AC}+\Omega_2^{AC}-\delta)t/2},
\end{dmath}
where the $\delta_{AC} = \Omega_1^{AC} - \Omega_2^{AC}$ is difference of AC Stark shift in levels $|1\rangle$ and $|2\rangle$. Here $\Omega_{eff}$ is the effective Rabi frequency for coupling of levels $|1\rangle$ and $|2\rangle$. The term $\phi_R$ is defined as $\phi_R = -{\bold{k_{eff}}}.\bold{r}+\omega_{eff}t+\phi_{eff}$, where $k_{eff}$ is the effective wave vector for two Raman beams and $\omega_{eff} = \omega_1-\omega_2$ and $\phi_{eff} = \phi_1-\phi_2$. Here the net coupling strength $\Omega_R$ is defined as $\Omega_R = \sqrt{\Omega_{eff}^2 - (\delta-\delta_{AC})^2}$. 

When an atom is initially ( at $t_{0}=0$ ) in level $|2\rangle$), then, after Raman beams pulse of duration $t$, the population remaining in level $|2\rangle$ is given as ,

\begin{equation}
   |C_{2}(t)|^{2} =  \left\{\cos^{2}\left(\frac{\Omega_R t}{2} \right)+\left(\frac{\delta-\delta_{AC}}{\Omega_R}\right)^{2}\sin^{2}\left(\frac{\Omega_R t}{2}\right)\right\}.
\end{equation}
The population (proportional to $|C_{2}(t)|^2$) in level $|2\rangle$ shows Rabi oscillations with variation in duration `t' of Raman beams. These oscillations strongly depend on the intensity of the Raman beams, detuning $\Delta$ from the excited state (level $|3\rangle$ ), and two-photon Raman detuning $\delta$. The experimetnally observed Rabi oscillations are described in section REF. The measurement of these Rabi oscillations are used to estimate the duration of $\pi$-pulse for Raman pulse atom interferometry.

\begin{figure}[h]
\begin{center}
\includegraphics[scale=0.45]{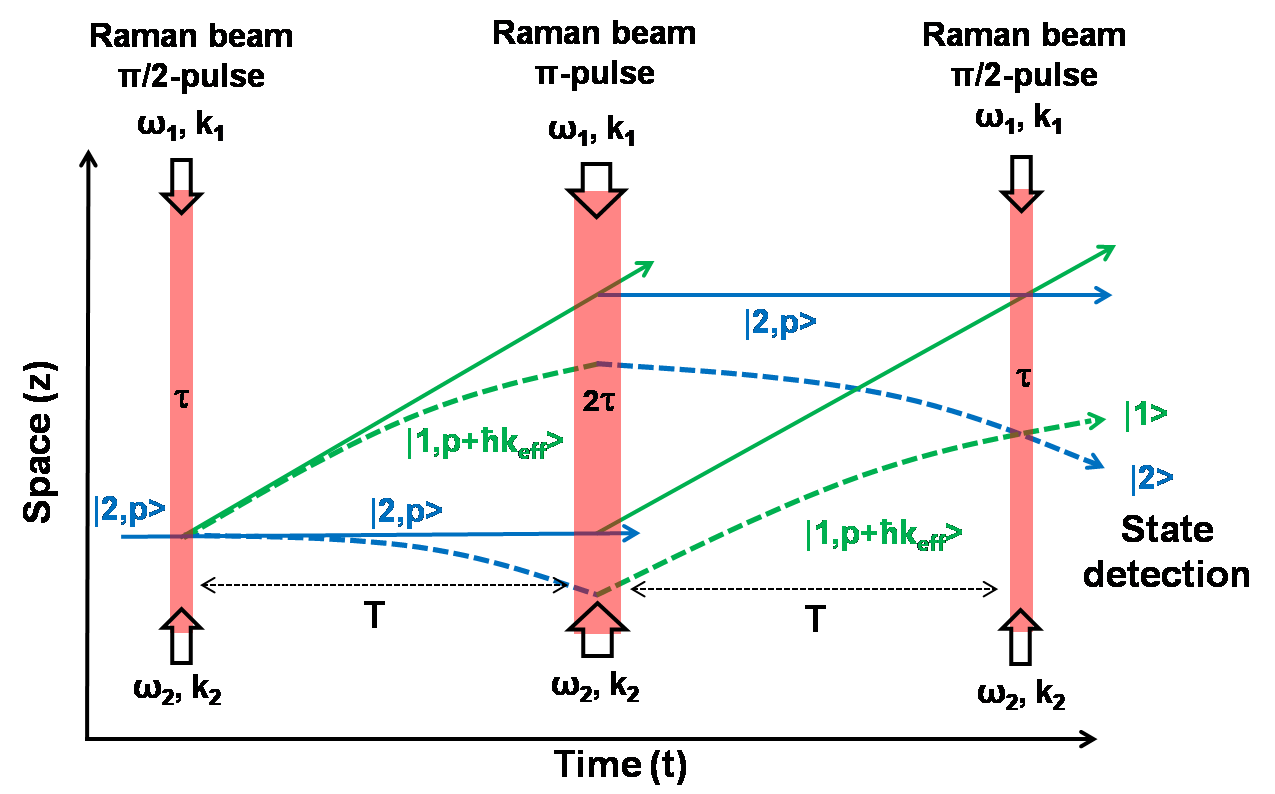}
\caption{ Schematic of the Mach-Zehnder type atom interferometer with three Raman light pulses sequence ($\pi/2 -T- \pi -T- \pi/2$). The first Raman laser pulse ($\pi/2$ pulse) creates a superposition of $|1\rangle$ and $|2\rangle$ states, which splits the atomic wave-packet in two different paths due to difference in momentum of states. The second pulse ($\pi$ pulse) inverts the states and reflects the atomic trajectories to recombine two paths. The third pulse ($\pi/2$ pulse) again creates a superposition of states which makes the two paths interfere.}
\label{MachZehnder}
\end{center}
\end{figure}

In the atom interferometry measurements, when atoms are moving under the influence of gravitational acceleration, the excitation of atoms between two levels  $|1\rangle$ and $|2\rangle$ is measured after applying a sequence of ($\pi/2$)-T-($\pi$)-T-($\pi$/2) pulses of counter-propagating Raman beams. The separation (T) between these pulses in the interferometer is kept much larger than the duration of the pulses. The impact of these Raman beams pulses on an atom in the atom cloud is to split, reflect and finally recombine the atomic wave-packet in this Mach-Zehnder type atom interferometer \cite{kasevich1992measurement}, as shown schematically in Fig. \ref{MachZehnder}. At the end of Raman pulses, the Raman transition probability from initial to final state is measured, which can give information about any quantum mechanical phase accumulated on atomic wave-packet during the Raman pulses in the interferometer. In absence of any external potential acting on the atom during the Raman pulses, i.e. no external quantum mechanical phase accumulated, the net Raman transition probability is zero due to ($\pi/2$)-T-($\pi$)-T-($\pi$/2) sequence of Raman pulses. This would result no change in populations in levels $|1\rangle$ and $|2\rangle$. This situation is depicted by paths with continuous lines in Fig. \ref{MachZehnder}. However, in presence of an external potential, such as gravitational potential, the accumulated phase in the interferometer is finite. In this situation, the paths in the interferometer are deviated from the original paths as shown by dashed curve in Fig. \ref{MachZehnder}. The finite value of phase accumulated in the interferometer would lead to modification in the net Raman excitation probability after the end of the Raman pulses. This would result in change in populations in levels $|1\rangle$ and $|2\rangle$ at the end of Raman pulses. When an atom is moving under the influence of gravitational acceleration (g), it experiences a continuously changing frequencies of Raman beams due to Doppler effect. This is reflected in the form of quantum mechanical phase accumulated in this atom interferometer, which is given as $\Delta \Phi = k_{eff}gT^{2}$, where $k_{eff} = |\va*{k_{1}}-\va*{k_{2}}| \approx 2k_{1}$ is the effective wave-vector magnitude of two Raman beams on the atom, and $\va*k_{1}$ and $\va*k_{2}$ are wave vectors of each Raman beam \cite{kasevich1992measurement,peters2001high}. Due to this phase, the change in the populations of levels $|1\rangle$ and $|2\rangle$ after the end of Raman pulses is observed, which is used for estimation of gravitational acceleration (g) \cite{kasevich1992measurement,peters2001high}. 

If we consider Raman excitation of N number of $^{87}Rb$ atoms from initial hyperfine state $|F = 2\rangle$ (i.e. state $|2\rangle$) to final hyperfine state $|F = 1\rangle$ (i.e. state $|1\rangle$) in the ground state of $^{87}Rb$ atom, the atomic populations in two hyperfine states $|F = 2\rangle$ and $|F = 1\rangle$, after the Raman excitation pulses ($\pi/2 -T- \pi -T- \pi/2$), are given as,
\begin{equation}\label{Prob_Phi2}
N_{|F=2\rangle} = \frac{N}{2}[1 + \cos(\Delta\Phi)] ,
\end{equation}
and
\begin{equation}\label{Prob_Phi1}
N_{|F=1\rangle} = \frac{N}{2}[1 - \cos(\Delta\Phi)] ,
\end{equation}
where $\Delta \Phi = k_{eff}gT^{2}$ is the total phase difference accumulated in the interferometer paths.

Thus, by measuring the populations in the hyperfine states $F = 1$ and $F = 2$, at the output of the interferometer, $\Delta \Phi$ can be known. From  $\Delta \Phi$, g can be estimated for a given T. A more convenient way to estimate g is to measure oscillations in the populations in any state, when frequency of one of the Raman beams is chirped during the Raman pulses sequence. This frequency chirp creates a time varying frequency difference between two Raman beams, which is applied to compensate the time varying Doppler shift of Raman beams frequencies experienced by a moving atom. In presence of frequency chirp rate ($\alpha$) applied to a Raman beam, the net phase in the interferometer is given as $\Delta \Phi = k_{eff}gT^{2}-2\pi\alpha T^{2} $. In the experiments, the population in a particular state can vary with the chirp rate. These oscillations in population are referred as interferometric fringes. The value of T affects the oscillation period in the fringes and hence affects the resolution of the measurements. For larger value of T, the fringes have smaller period of oscillations with increased sharpness. 

After the Raman pulses, if the population is measured in the initial state (i.e. $|F = 2\rangle$), the net phase in the interferometer is zero (or multiple of 2$\pi$) for a peak in the fringes. At a particular chirp rate, the fringes observed for different values of T have a common maximum in the fringes. For this chirp rate, called central chirp rate ($\alpha_{c}$), the net phase in the interferometer is zero. This can be written as, 
\begin{equation}\label{total_Phi}
\Delta \Phi = k_{eff}gT^{2}-2\pi\alpha_{c}T^{2} = 0,
\end{equation} 
which gives $g = (2\pi \alpha_{c})/k_{eff} = \alpha_{c} \lambda/2$. Here $k_{eff} = 2k_{1} = 4\pi/\lambda$, where $\lambda$ is wavelength of Raman beam. Thus, by measuring fringes for different values of T, one can find the common fringe peak and the chirp rate $\alpha_{c}$. From the value of $\alpha_{c}$, g can be estimated.

\section{Experimental system}
CAG presented here is based on Doppler sensitive two-photon Raman pulse atom interferometry technique. The measurements are performed when cold $^{87}Rb$ atoms are launched in a fountain geometry and atoms are moving against gravity. The setup is consisting of a magneto-optical trap (MOT) for $^{87}Rb$ atoms, a mechanism for vertical launching of cold atoms in atomic fountain and Raman beams for interferometric excitation of atoms in the fountain. All these functions are performed by a single laser system whose output is divided in several laser beams and their frequencies are adjusted using different modulators. For the state detection, after interferometry, absorption probe beam was derived from a different low power laser.

\subsection{Magneto-optical trap} 

\begin{figure*}[h]
\centering
    \begin{subfigure}[t]{0.5\textwidth}
    \centering
   \includegraphics[scale=0.5]{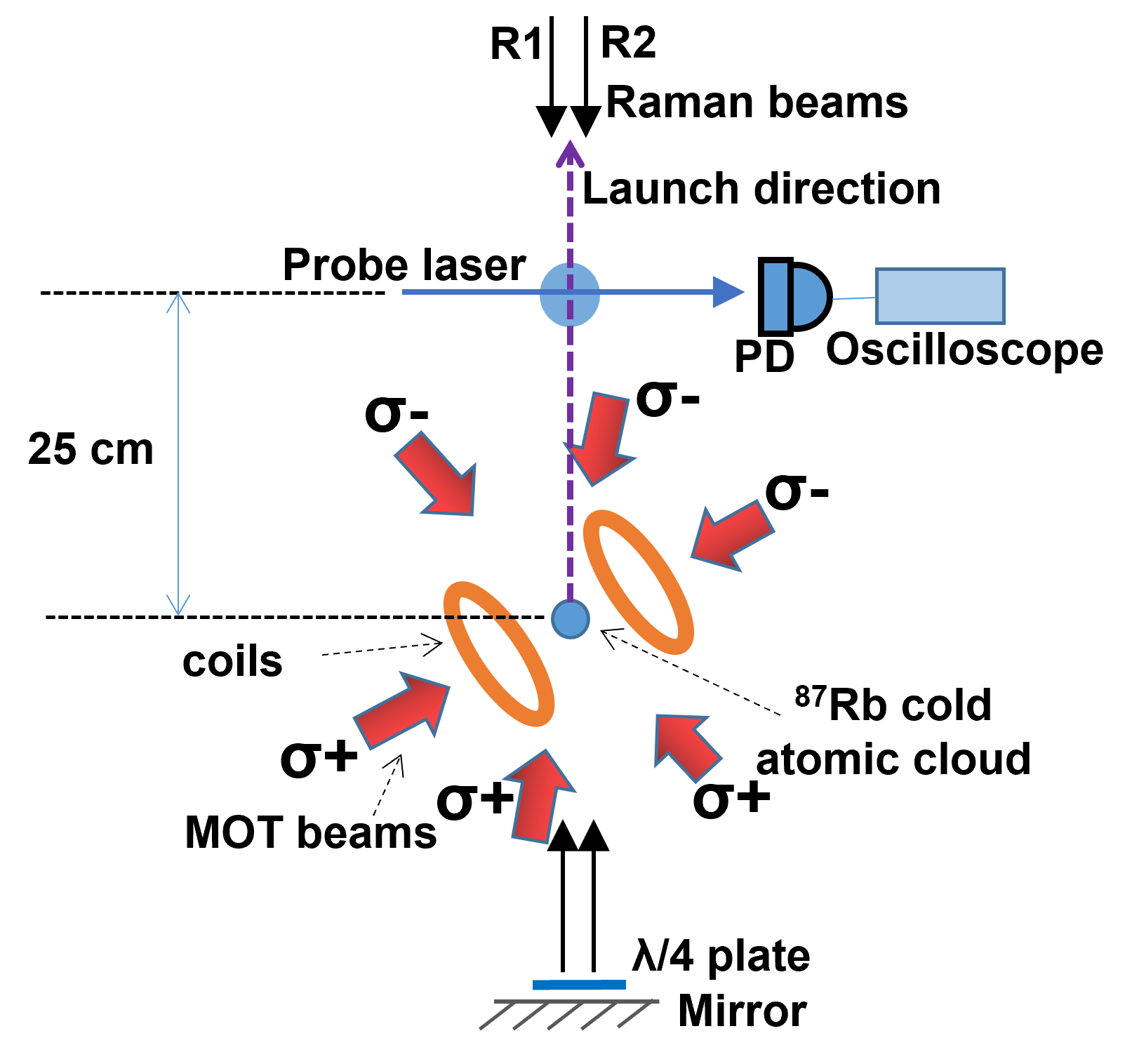}
\caption{}
\label{FountainScheme}
\end{subfigure}\hfill
\begin{subfigure}[t]{0.5\textwidth}
    \centering
    \includegraphics[width=\textwidth, height=0.75\linewidth,keepaspectratio]{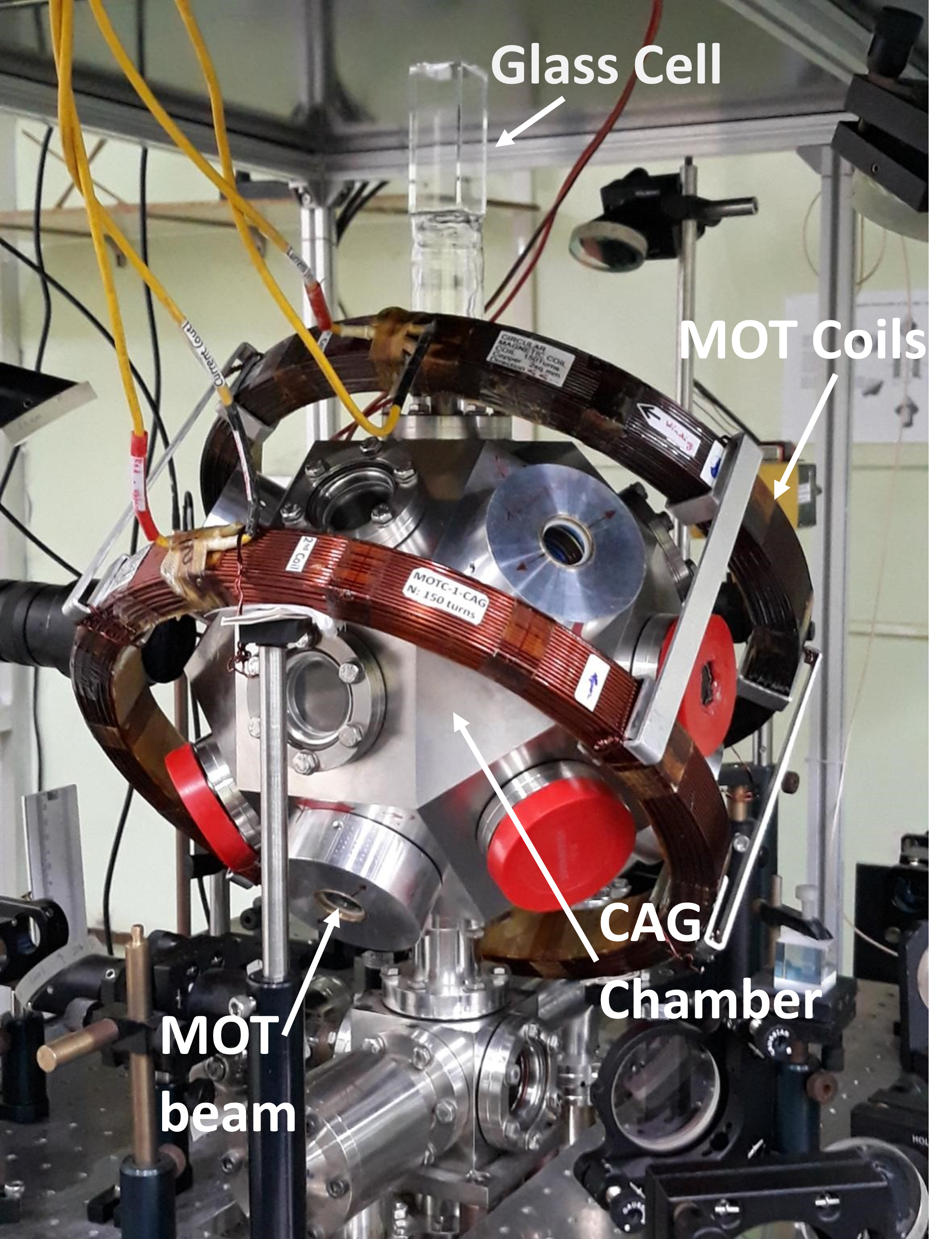}
    \caption{}
    \label{Setup}
\end{subfigure}\hfill
    \caption{(a) MOT beams, Raman beams and probe beam in the experimental setup. (b) Photograph of the setup.}
\end{figure*}

The cold $^{87}Rb$ atoms used in this gravimeter (CAG) setup are trapped in magneto-optical trap loaded from background Rb-vapor in the chamber. The cooling laser beams in this MOT (MOT beams) are aligned in (1,1,1) configuration, in which vertical direction is kept as (1,1,1) direction in the frame of coordinate axes along the MOT beams (Fig. \ref{FountainScheme}). In this configuration, since there is no MOT beam in the vertical direction, the MOT beams do not hinder the path of atoms when they are launched in fountain in vertical direction for atom interferometry measurements of g.  We have derived all laser beams from the output of a single extended cavity diode laser (ECDL) system (DLC-TA-Pro, TOPTICA, Munich, Germany) operating at a wavelength of 780 nm. This makes our gravimeter system compact and easy to use \cite{singh2021single}. 

\begin{figure*}[h]
\centering

\begin{subfigure}[t]{0.5\textwidth}
    \centering
    \includegraphics[scale=0.41]{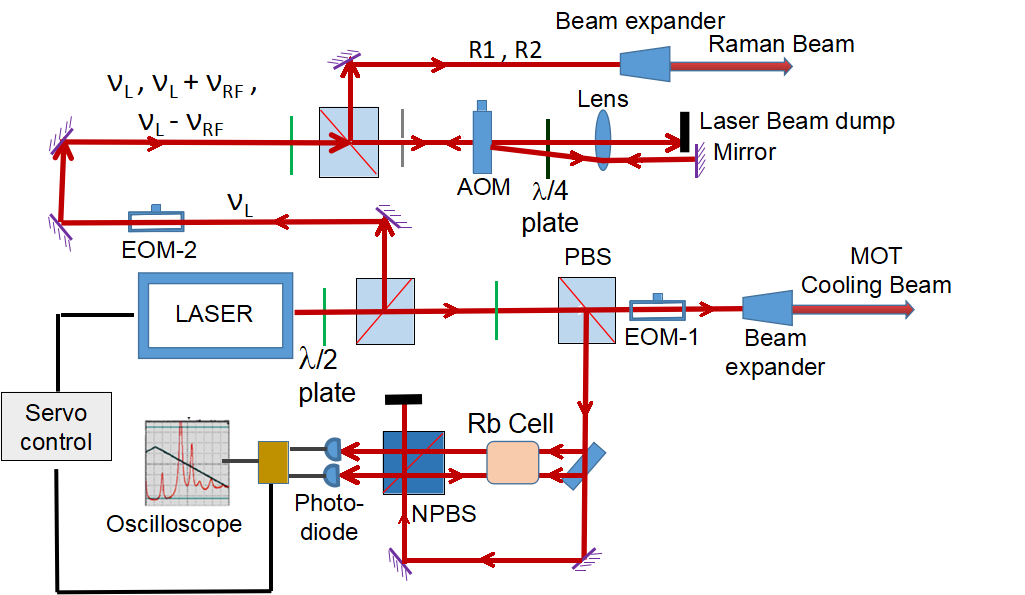}
    \caption{}
    \label{Raman Beam Generation}
\end{subfigure}\hfill
\begin{subfigure}[t]{0.4\textwidth}
    \centering
    \includegraphics[scale=0.37]{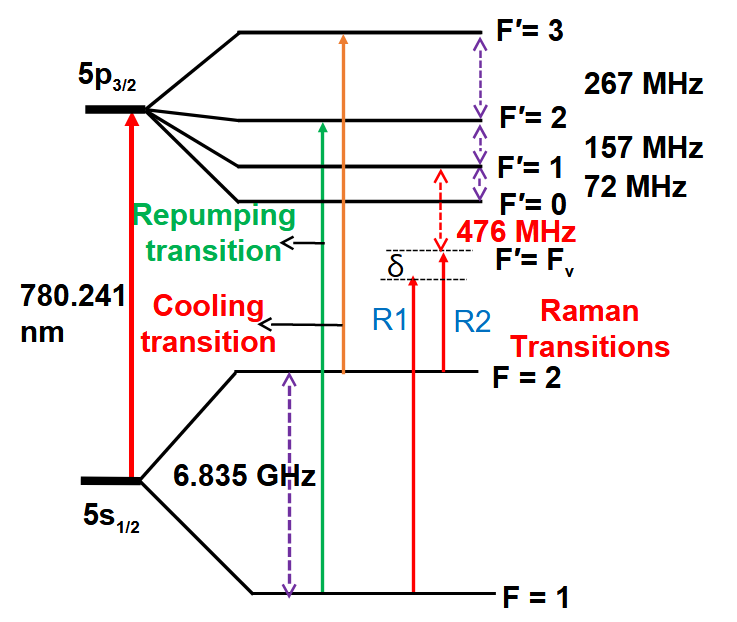}
    \caption{}
    \label{EnergyLevels}
\end{subfigure}\hfill

    \caption{(a) Schematic of optical layout for Raman beams and MOT beam generation for cold atom gravimeter (CAG), (b) The schematic of energy levels of $^{87}Rb$ involved in cooling, repumping and Raman transitions  of CAG setup.}
\end{figure*}

The cooling and re-pumping laser emissions for $^{87}Rb$ atom are generated by passing a part of output of the ECDL through an electro-optic phase modulator (EOM-1) operated at 6.58 GHz, as shown in Fig. \ref{Raman Beam Generation}. The output laser beam from EOM-1, having cooling and re-pumping emissions, is split into two beams which are passed through two independent acousto-optic modulators (AOMs) to control frequency and power in output beams. Output of one AOM is split into three beams, to prepare three MOT beams which are directed to UHV chamber from lower side of the chamber. The output beam of the other AOM is split into other three MOT beams which are directed to UHV chamber from the upper side of the chamber. The intensity of each of these six MOT beams is kept $\sim$ 25 $mW/cm^{2}$. The different MOT beams enter into the ultra-high vacuum (UHV) chamber from different view ports on the chamber. The MOT beams intersect one another at the centre of the UHV chamber where zero of the quadrupole magnetic field is set. The frequency of the cooling beams was kept $\sim$12 MHz red-detuned with respect to the cooling transition $ 5 $ $^{2}S_{1/2} (F = 2)$ $\rightarrow$ $ 5 $ $^{2}P_{3/2} (F^{'}= 3)$ of $^{87}Rb$. The second part of the output from ECDL is passed through another electro-optic modulator (EOM-2) to generate two Raman beams ($R_{1}$ and $R_{2}$) at different frequencies for atom interferometry purpose. The schematic of the generation of MOT beams and Raman beams for the CAG setup is shown in Fig. \ref{Raman Beam Generation}. The ECDL (DL-Pro) is frequency locked by saturated absorption spectroscopic technique. The relevant transitions of $^{87}Rb$ used for cooling, repumping, probing and Raman transitions are shown in Fig. \ref{EnergyLevels}. 

A pair of coils connected in quadrupole configuration is used to generate quadrupole magnetic field for MOT. A direct current (DC) of 11.5 A flowing in these coils generates the desired field gradient of $\sim$ 10 G/cm for MOT. The Rb vapor is injected into the chamber by passing a DC current of 3.8 A in Rb-dispenser source (SAES Rb getter). Initially, $^{87}Rb$ atoms are cooled and trapped in the MOT. The atom cloud in MOT has typically 2.5 x $ 10^7 $ number of atoms at temperature of $\sim$ 70 $\mu$K. For further lowering of temperature of atom cloud, we implement the optical molasses stage. \\

\subsection{Atomic fountain}

For launching of atoms in vertically upward direction in atomic fountain geometry, the frequency of three cooling laser beams propagating in downwards direction was shifted towards red side with respect to the frequency of the other three cooling beams \cite{singh2021single}. The launching velocity of atoms in (1,1,1) configuration is given as \cite{Zhou}, $v_{launch}$ = $\sqrt{3}$$\lambda \Delta \nu$/2, where $\Delta \nu$ is the frequency difference between upward and downward propagating cooling laser beams in MOT and $\lambda$ is the wavelength of cooling beam. For the detection of launched atoms at a particular height ($\sim$25 cm) from the centre of the MOT, a horizontal resonant probe laser beams as shown in Fig. \ref{FountainScheme} is used. This probe beam has $\sim5\mu$W power and 1 mm size ($1/e^2$ radius). The probe absorption signal is highly dependent on the initial temperature of the atom cloud in the molasses. The lower temperature of the cloud allows more atoms reaching the probe beam before their loss in transverse direction. We have observed that atom cloud temperature in molasses is effectively reduced by switching-off the MOT coil current before the beginning of the molasses stage in which detuning of the cooling beams is increased. So we switched-off MOT coils current 30 ms before molasses stage, which results in reduction in temperature to $\sim$ 25 $\mu$K with $1.8 \times 10^7 $ number of atoms in the cloud. After launching this lower temperature cloud, a considerable increase ($\sim$ 5 times at 9 cm height) in the atomic flux in the fountain was observed.\\
 
\begin{figure}[h]
\centering
    \includegraphics[scale=0.35]{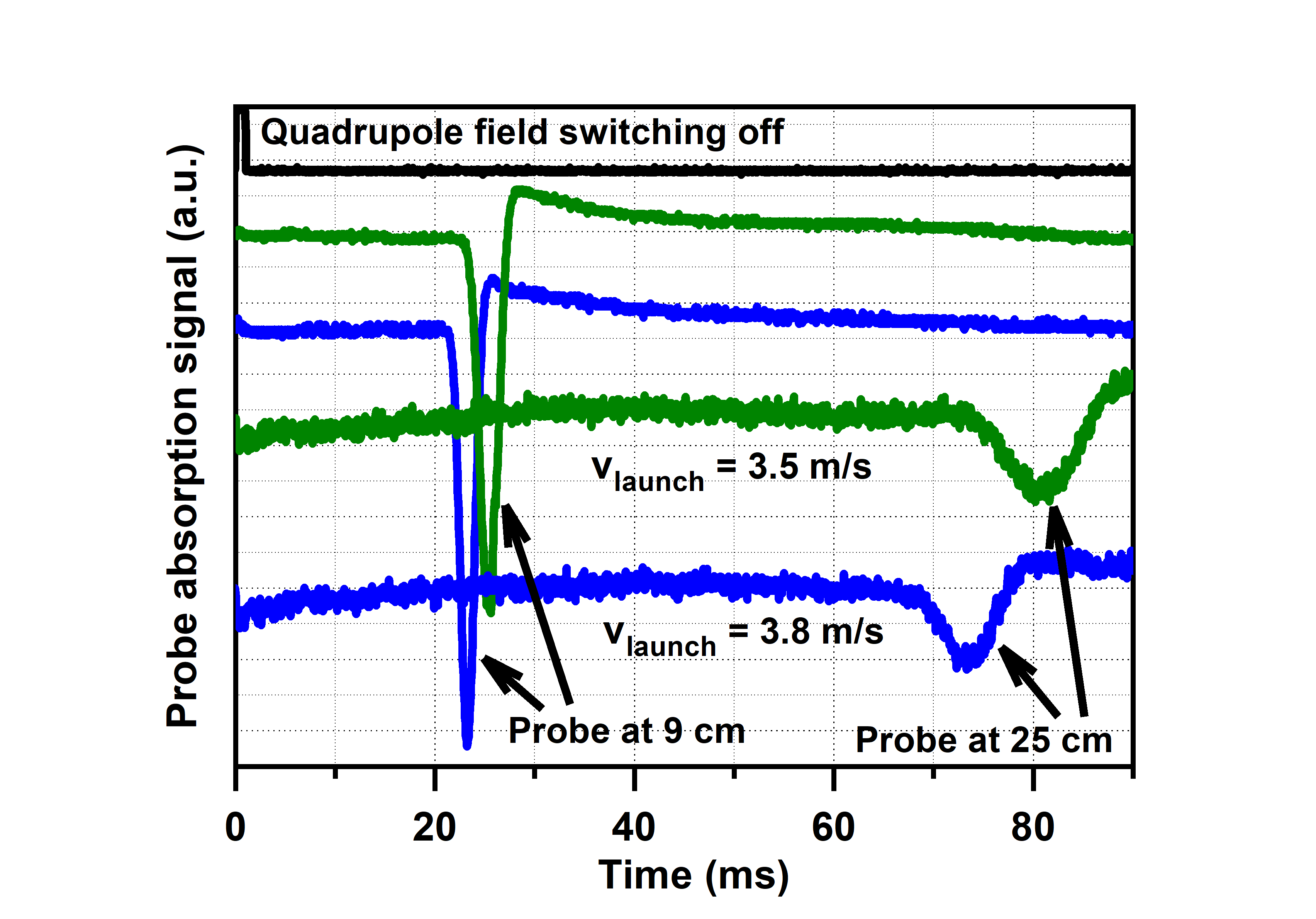}
    \caption{Observed probe absorption signal at 9 cm and 25 cm height in fountain for different launch speeds (blue :3.8 m/s and green : 3.5 m/s) of atom cloud.}
    \label{Flux Absorption probe}
\end{figure}

After the launch of atom cloud, we have monitored the atom flux at different heights by recording an absorption signal from an horizontal absorption probe beam. When atoms in fountain crossed this beam probe, the absorption of probe was recorded. The probe absorption signal at heights of 9 cm and 25 cm are shown in Fig. \ref{Flux Absorption probe}. This figure also shows the signals for different values inital launch speed of atoms. Green and blue curves are for launch speed of 3.5 and 3.8 m/s. It was also verified that the theoretically calculated launch speed was in good agreement with the speed estimated from time of flight recorded in the absorption signal.

\subsection{State selection} 
 
We have prepared $^{87}Rb$  atoms in $|F = 2\rangle$ hyperfine level of ground state. By virtue of stray magnetic field which is $\sim$ 500 mG, the degeneracy of the magnetic sublevels in $|F = 1\rangle$ and $|F = 2\rangle$ states is removed. We tuned the frequency of Raman beams such that the Raman transition $|F = 2, m_{F} = 0\rangle$ $\rightarrow$ $|F = 1, m_{F} = 0\rangle$ is excited. In order to verify this, we applied a $\pi$-pulse of Raman beams on the atom cloud (in $|F = 2\rangle$ state), after $\sim$ 5 ms of its launch in the fountain, and recorded the absorption signal of a probe beam at 9 cm height in the fountain path. The frequency of this probe was tuned to $|F = 1\rangle$ $\rightarrow$ $|F^{'} = 0\rangle$ transition. We observed that probe absorption was significant when probe beam was linearly polarized which was corresponding to $\pi$ polarization of the probe beam for $|F = 1, m_{F} = 0\rangle$ $\rightarrow$ $|F^{'} = 0, m_{F^{'}} = 0\rangle$ transition. There was negligible absorption of probe, when probe beam polarization was made circular polarization (left handed and right handed). A fine tuning of probe beam frequency also did not alter these results. Therefore, these results indicate the negligible population transfer to states $|F = 1, m_{F} = -1\rangle$ and $|F = 1, m_{F} = 1\rangle$ due to applied $\pi$-pulse of Raman beams. Thus our Raman beams excite atoms between $|F = 2, m_{F} = 0\rangle$ $\rightarrow$ $|F = 1, m_{F} = 0\rangle$ states.       
\subsection{Raman pulse atom interferometer} 
 
In the experiments of Raman pulse atom interferometry for g measurements, we have taken two Raman beams $R_1$ \& $R_2$ having power ratio of 3:1 (total Raman beam power = 16 mW) and beam size of approximately 1.2 cm. To generate Raman beams, the ECDL laser output beam was locked at a frequency ($\nu_{L}$) that is $\sim$ 12 MHz blue-detuned from $C_{13}$ crossover transition (corresponding to $| 5 $$^{2}S_{1/2} F = 2 \rangle $ $\rightarrow$ $| 5 ^{2}P_{3/2} F^{'}= 1\rangle$ and $| 5 $$^{2}S_{1/2} F = 2\rangle$ $\rightarrow$ $| 5 $ $^{2}P_{3/2} F^{'}= 3\rangle$ of $^{87}Rb$). This laser beam was passed through an electro-optic phase modulator (EOM-2 in Fig. \ref{Raman Beam Generation}) working at $\nu_{RF} \sim$ 6.83 GHz giving output at three frequencies $\nu_{L}, \nu_{L} +\nu_{RF}$ and $\nu_{L} - \nu_{RF}$. The power in each of the side-band is $\sim$ 20$\%$ of the  the transmitted power. The side-band at frequency $\nu_{L} - \nu_{RF}$ remains unutilized in the experiments. The output of this EOM was double-passed through an acousto-optic modulator (AOM) working at 350 MHz. The emission in two beams at frequencies $\nu_{2}$ = $\nu_{L}$ - 700 MHz and $\nu_{1}$ = $\nu_{L} +\nu_{RF}$ - 700 MHz from AOM served respectively as Raman beams $R_2$ and $R_1$ in our experiments with angular frequencies as $\omega_{2}$ = 2$\pi$$\nu_{2}$ and $\omega_{1}$ = 2$\pi$$\nu_{1}$. Thus, the Raman beam $R_2$ is red detuned by $\sim$ 476 MHz from $| 5 $ $^{2}S_{1/2} F = 2 \rangle$ $\rightarrow$ $| 5 $ $^{2}P_{3/2} F^{'}= 1 \rangle$ transition. Both the Raman beams, $R_1$ and $R_2$, are used in retro-reflected geometry where polarisation of the beams are rotated by 90 degree by keeping a quarter-wave plate before retro-reflecting mirror as shown in Fig. \ref{FountainScheme}. The chirping or extra detuning of the Raman beams was implemented through EOM-2 (Fig. \ref{Raman Beam Generation}), which shifted the frequency of Raman beam $R_1$ at frequency $\nu_{1}$. In our setup, since we tuned our Raman beams such that the upward going Raman beam R2 and downward going Raman beam R1 interact with the atoms, the frequency of R1 needs to be chirped-up for keeping the two photon resonance with the reducing speed of atoms due to g. \\

\begin{figure*}[h]
\centering
    \centering
    \includegraphics[scale = 0.35]{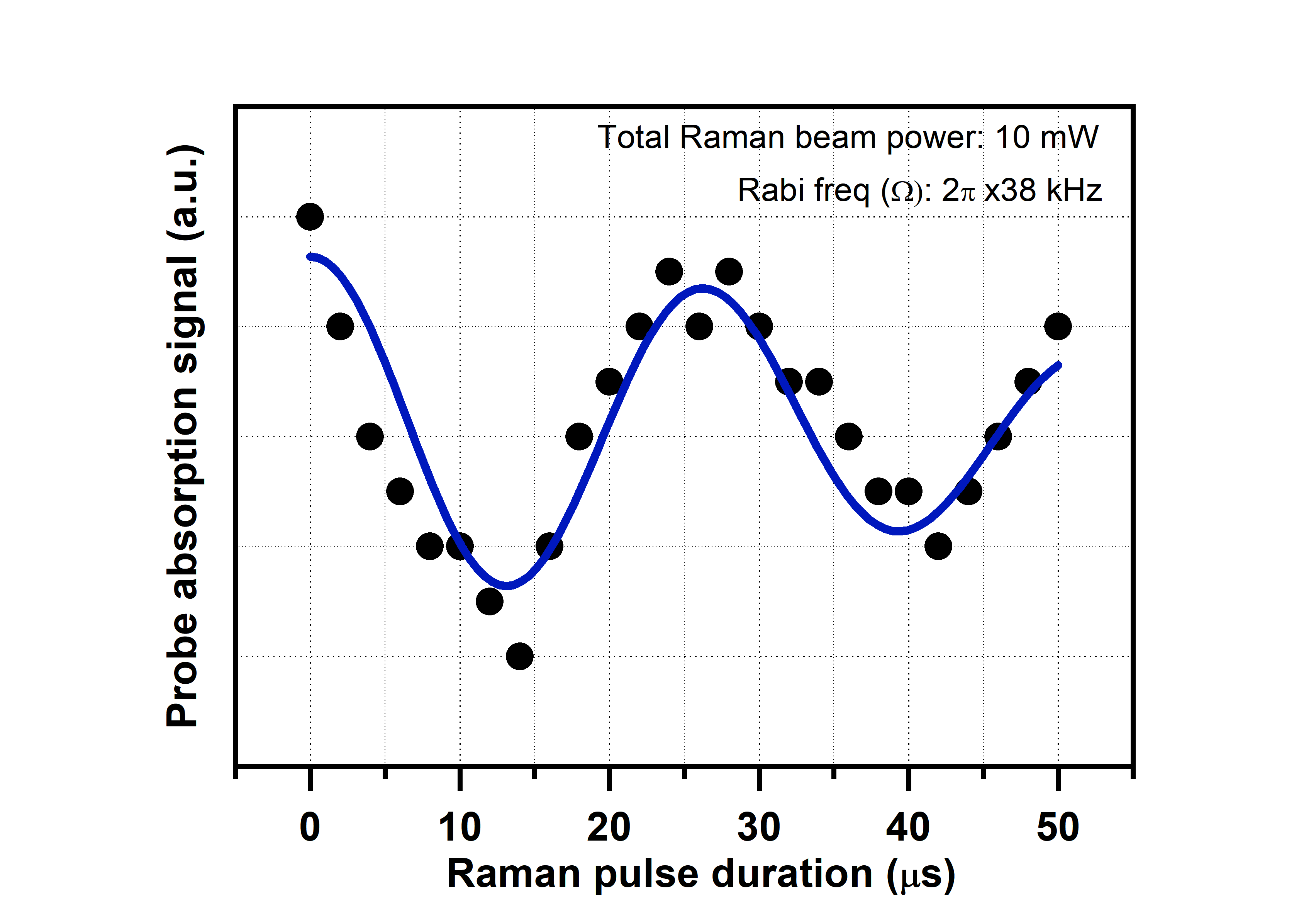}
    \label{Rabi Oscillation 10 mW}
    \caption{Rabi oscillations in atomic population in $|F = 2\rangle$ hyperfine ground state of $^{87}{Rb}$ for Raman beam power of 10 mW. The circles show measured data and continuous curve is guide to the eye for the measured data.}
    \label{Rabi Oscillation}
\end{figure*}

In order to study the Rabi oscillations under two-photon Raman transitions between two ground hyperfine states of $^{87}Rb$, which is required to estimate the duration of $\pi$-pulse for atom interferometry, the Raman beams (with total power of 10 mW in R1 and R2) are switched-on after 5 ms of launch of atoms in the fountain. The atoms from molasses are launched in their state $|5 ^{2}S_{1/2} F = 2\rangle$. The atoms state detection is done by applying an absorption probe at height of 25 cm from the MOT centre. By changing the duration of Raman beams, the absorption signal (proportional to number of atoms in hyperfine level F=2 of $^{87}$Rb) was recorded. These results are shown in Fig. \ref{Rabi Oscillation}. As observed in this figure, the amplitude of the oscillations damps as the Raman beams pulse duration is increased. This damping can be associated to the degradation of the coherence as well as reduction in number of atoms due to thermal expansion of the cloud in the atomic fountain. The $\pi$-pulse duration for Raman beams can be estimated for maximum transfer of cold $^{87}$Rb atoms from  $|5 ^{2}S_{1/2} F = 2\rangle$ to $|5 ^{2}S_{1/2} F = 1\rangle$, and this is $\sim$ 13.25 $\mu$s at 10 mW power. The power indicated here is total power in both the beams with ratio of power distribution 3:1. The Rabi frequency experimentally estimated from the data is  2$\pi \times$ 38 kHz which is close to calculated theoretical value of Rabi frequency 2$\pi \times$ 40 kHz at total Raman beam power of 10 mW (calculated using Eq. \ref{SRT_Multiple}).

\subsection{SCADA system for Cold atom gravimeter setup} 

In cold atom gravimeter setup, various events from MOT loading to detection of interference fringes in the atom interferometer are controlled by in-house developed Supervisory Control and Data Acquisition (SCADA) system. This system controls the sequence of operation of various devices such as acousto-optic modulators (AOMs), power supplies, etc., in a time synchronised manner. The CAG setup needs a number of control signals to control various devices. The controller system can provide more than 50 control signals to perform various steps in the experiments. The control signals are used in a series of steps to perform different functions such as formation of MOT, launching of atoms in fountain, Raman pulse atom interferometry, detection of atoms in a particular state, etc. In each step, a number of instruments are used to execute various tasks. The duration for which a particular step is executed with the synchronisation between instruments, the programmed control signals are critical for the successful completion of each task. For these control signals, one typically needs temporal resolution of the order of tens of micro-seconds. This precise timing requirement is difficult to achieve with processors. To meet requirement of stringent time synchronization between various analog and digital control signals, a Field Programmable Gate Array (FPGA) based controller is selected to generate the control signals and perform various tasks in the gravimeter operation. To achieve temporal resolution of 1 $\mu$s, a 40 MHz reference clock is used for synchronization of events. Timing synchronization between control signals is of the order of 1 $\mu$s. To obtain precise control drive signal to AOMs and magnetic coil power supplies, all analog control signals have 16-bit resolution. 

A software framework is designed for performing the experiment which gives flexibility to generate control signals as per requirement.  This framework ensures that the FPGA is automatically re-configured to generate every desired pattern of the control signals. A software for personal computer (PC) is also developed which provides intuitive Graphical User Interface (GUI) for configuring the experimental parameters. There are 32 analog control signal (out of which 4 are isolated) to control AOMs, power supplies, etc. There are 24 digital signals to control various devices such as CCD camera, power supplies, shutters, etc. This controller is developed on NI make programmable automation controller c-RIO 9035. 

To estimate the sensitivity of `g' measurements with our CAG setup, the measurement cycle with updated parameters (such as chirp rate) needs to be repeated several hundred times (typically 500-600 times), in a stable ambient environment. This process is manually difficult and time consuming. Therefore, an automation of ‘g’ measurement is carried out. For this purpose, data acquisition of atom interferometer signal from Keysight DSOX 2024A is done for variation in the chirp rate value (achieved by the arbitrary wave function generator Tektronix AFG3102C). A computational algorithm is implemented in software to compute the value of ‘g’ from the acquired data set. The GUI finally displays the computed value of earth's gravitational acceleration `g'.

\section {Results and discussion}

\subsection {Interference fringes}
\begin{figure}[h]
\begin{center}
\includegraphics[scale=0.31]{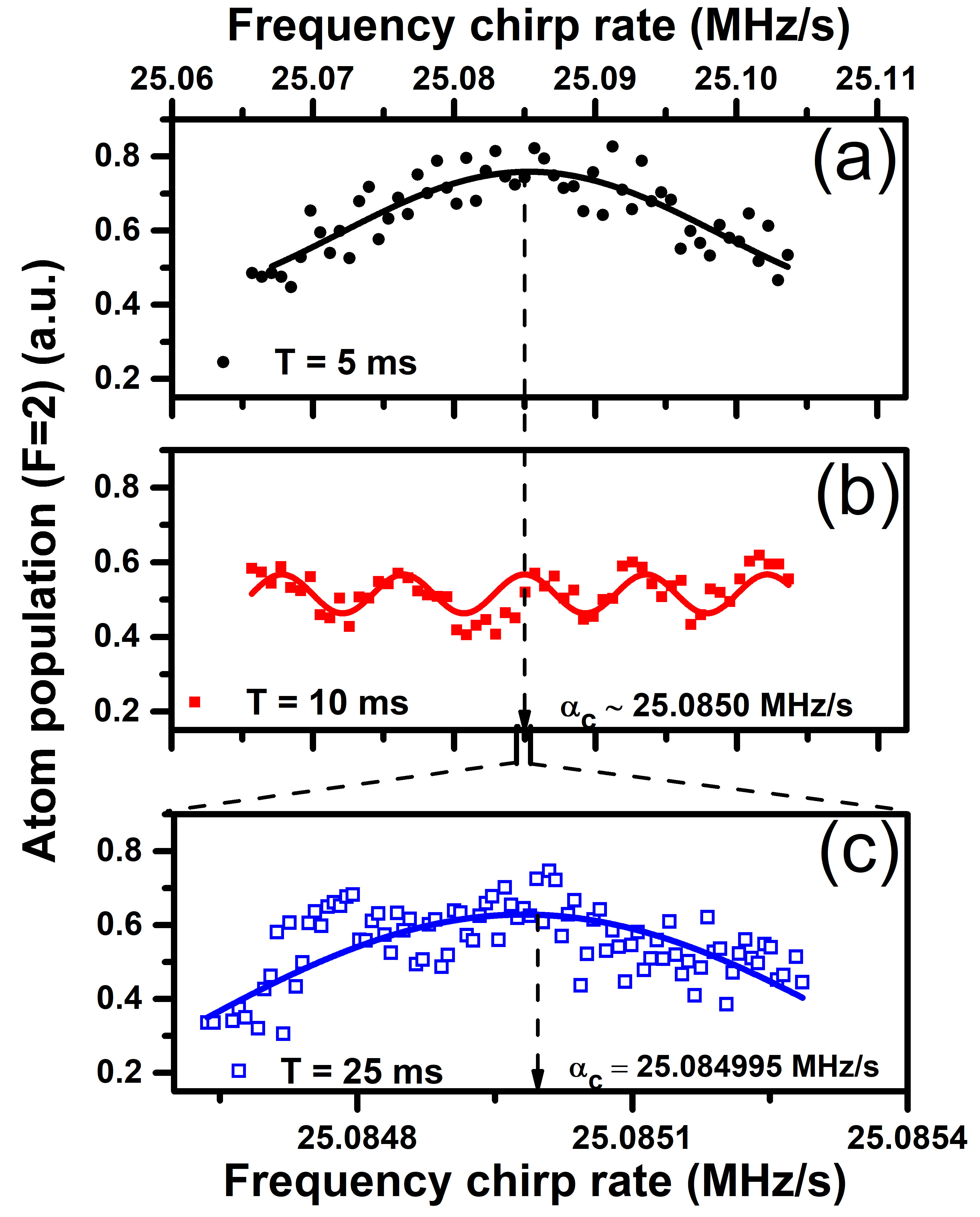}
\caption{The interference fringes with Raman beam frequency chirp rate ($\alpha$). Plot (a) shows fringes for T = 5 ms (circles).  Plot (b) shows the fringes for T = 10 ms (squares), at the same chirp rate scale as plot (a). Plot (c) shows the expanded view of the fringe at $\alpha_{c}$ recorded for T = 25 ms (hollow squares). In plot (c), the minimum step in frequency chirp rate was 6.9 Hz/s which corresponds to 270 $\mu$Gal in terms of acceleration.}
\label{Fringe_combined}
\end{center}
\end{figure}

After launching the cold $^{87}{Rb}$ atom cloud ($\sim$ 25 $\mu$K) in vertically upward direction and applying Raman pulses ($\pi/2 -T- \pi -T- \pi/2$), the change in population in $|F = 2\rangle$ state was monitored. The total power in Raman beams (R1 and R2) was 16 mW with power distribution ratio of 3:1 in two beams. The corresponding $\pi$ pulse duration was 10 $\mu$s. In these atom interferometry measurements, the population in $|F = 2\rangle$ state was measured for different values of chirp rate ($\alpha$). The measurements were repeated by changing the value of T in sequence $\pi/2 -T- \pi -T- \pi/2$ of Raman pulses. The Raman excitation pulses were applied during the upward motion of the atom cloud. The results of oscillations in population (in $|F = 2\rangle$ state) with chirp rate ($\alpha$) are shown in Fig. \ref{Fringe_combined}. These oscillations are commonly referred as intereference fringes.

Out of two Raman beams in retro-reflection arrangements, the frequency of Raman beam R1 was chirped at a chirp rate $\alpha $, to compensate the changing Doppler shift in two-photon resonance due to changing speed of atoms in presence of gravitational acceleration (g). In this arrangement, out of two pairs of counter-propagating beams, only one pair of counter-propagating beams, i.e. beam of frequency $\omega_{1}$ propagating in downward direction and beam of frequency $\omega_{2}$ propagating in upward direction, results in population transfer from initial state to final state. In order to know the population of atoms in a particular state $|F = 2\rangle$ after interferometry sequence, the absorption signal of probe beam, aligned at 25 cm height from the MOT centre, was monitored.

Here in Fig. \ref{Fringe_combined}c, the plot shows the results of measurement of interference fringes with an increased resolution of 6.9 Hz/s in the chirp rate. This resolution of 6.9 Hz/s in chirp rate corresponds to 270 $\mu$Gal resolution in terms of acceleration (g). Fig. \ref{Fringe_combined} shows the interferometric fringes taken without seismic vibration correction \cite{vibrationcorrection} for different values of T (5 ms, 10 ms and 25 ms). The common peak in the fringes for all values of T gives the central chirp rate chirp rate $\alpha_{c}$ for which net interferometric phase is zero. Using the obtained value of $\alpha_{c}$ = 25.084995 MHz/s, we obtain the local value of ‘g’ in our laboratory to be g = 9.786173 $m/s^{2}$. This value is in agreement with the value of g for Indore city calculated using Earth's Gravitational Model 2008 \cite{egm2008}.

\begin{figure}[h]
\begin{center}
\includegraphics[scale=0.35]{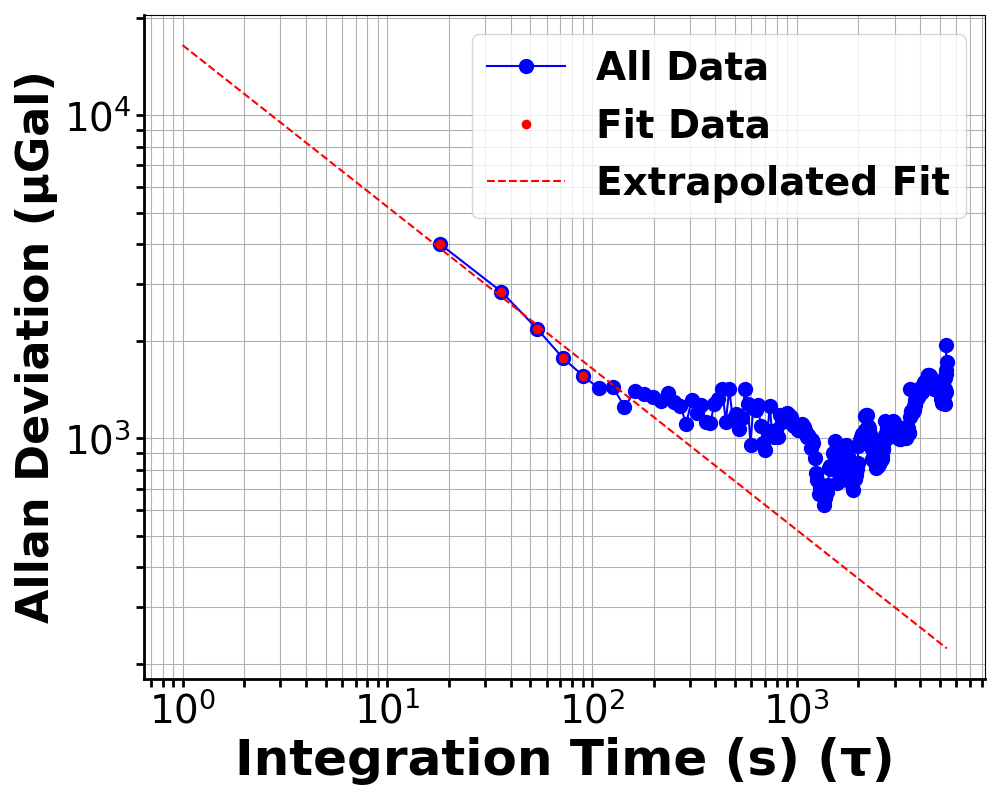}
\caption{Measured Allan deviation in acceleration due to gravity for
the Raman pulse separation of T = 25 ms.}
\label{fig:allan_dev}
\end{center}
\end{figure}
\subsection {Allan variance}
 In order to estimate the sensitivity of our measurements, we have estimated the Allan deviation in the measurement of g. The Allan deviation, which is square root of Allan variance, is a commonly used parameter to characterize the temporal noise sensitivity in the precision measurements . For this, data of measured g value was recorded over a large number (N=600) of cycles of interferometric measurement, at a fixed value of $\alpha$ and T=25 ms. Each cycle of measurement in our experiments take nearly 18 seconds. From these measurements, the Allan variance, was calculated using the relation, 
 \begin{equation}
\sigma^2(\tau) = \frac{1}{2\left( M − 1 \right)} \sum_{i}^{M-1} \left( g_{i+1}\left(\tau\right) - g_{i}\left(\tau\right) \right)^2,
 \end{equation} 
 where M is the number of data points $g_{i}$($\tau$) each recorded at integration time of $\tau$. The results of measurements of Allan deviation are shown in Fig. \ref{fig:allan_dev} for different values of integration time ($\tau$). Here, the dashed line shows a fit to the measured Allan deviation at smaller $\tau$, corresponding to white noise dominated behaviour of Allan deviation ($\sigma$ $\propto$ $\tau^{−1/2}$). From the graph in Fig. \ref{fig:allan_dev}, it is obvious that measured Allan deviation initially reduces with $\tau$ and reaches to minimum value of 621 $\mu$Gal at integration time $\tau$ = 1350 s. With further increase in $\tau$, the Allan deviation increases, which could be attributed to other noises in the setup such as drift and random walk.  
 
\section*{Conclusion}
We have developed a cold atom gravimeter setup in atomic fountain geometry using a single laser system. In the setup, the laser cooled $^{87}Rb$ atoms are launched in the fountain geometry and two-photon Raman pulse atom interferometry is used to measure the gravitational acceleration acting on the atoms. The local value of acceleration due to gravity has been estimated with a sensitivity of 621 $\mu$Gal for integration time of 1350 s.  

\section*{Acknowledgments} We are thankful to Sanjeev Bhardwaj for the help during the experiments. We acknowledge the contribution of Hemraj Bundel, Lalita Jain, P. P. Deshpande and V. P. Bhanage in the development of SCADA controller system. We are also thankful to S. V. Nakhe for his keen interest in the work and continuous encouragement. 
\\\\

\begin{thebibliography}{10}

\bibitem{inguscio}
M.~Inguscio and L.~Fallani, ``Atomic physics: Precise measurements and ultracold matter,'' {\em ISBN: 9780198525844}, 2013.

\bibitem{BEC}
S.~R. Mishra, S.~P.~P. Ram, S.~K. Tiwari, and H.~S. Rawat, ``Dependence of in-situ bose condensate size on final frequency of rf-field in evaporative cooling,'' {\em Pramana}, vol.~88, pp.~1--11, 2017.

\bibitem{Fermi}
I.~Bloch, J.~Dalibard, and S.~Nascimb{\`e}ne, ``Quantum simulations with ultracold quantum gases,'' {\em Nature Physics}, vol.~8, no.~4, pp.~267--276, 2012.

\bibitem{Kasevich1999}
P.~Asenbaum, C.~Overstreet, M.~Kim, J.~Curti, and M.~A. Kasevich, ``Atom-interferometric test of the equivalence principle at the ${10}^{\ensuremath{-}12}$ level,'' {\em Phys. Rev. Lett.}, vol.~125, p.~191101, Nov 2020.

\bibitem{FineStructure}
P.~Clad\'e, E.~de~Mirandes, M.~Cadoret, S.~Guellati-Kh\'elifa, C.~Schwob, F.~m.~c. Nez, L.~Julien, and F.~m.~c. Biraben, ``Determination of the fine structure constant based on bloch oscillations of ultracold atoms in a vertical optical lattice,'' {\em Phys. Rev. Lett.}, vol.~96, p.~033001, Jan 2006.

\bibitem{rosi2014precision}
G.~Rosi, F.~Sorrentino, L.~Cacciapuoti, M.~Prevedelli, and G.~Tino, ``Precision measurement of the newtonian gravitational constant using cold atoms,'' {\em Nature}, vol.~510, no.~7506, pp.~518--521, 2014.

\bibitem{Tino}
G.~M. Tino and F.~Vetrano, ``Is it possible to detect gravitational waves with atom interferometers?,'' {\em Classical and Quantum Gravity}, vol.~24, p.~2167, apr 2007.

\bibitem{atomic}
P.~Arora, A.~Awasthi, V.~Bharath, A.~Acharya, S.~Yadav, A.~Agarwal, and A.~S. Gupta, ``Atomic clocks: A brief history and current status of research in india,'' {\em Pramana}, vol.~82, pp.~173--183, 2014.

\bibitem{inertial}
R.~Geiger, A.~Landragin, S.~Merlet, and F.~Pereira Dos~Santos, ``{High-accuracy inertial measurements with cold-atom sensors},'' {\em AVS Quantum Science}, vol.~2, 06 2020.
\newblock 024702.

\bibitem{car-gravimeter}
J.-Y. Zhang, W.-J. Xu, S.-D. Sun, Y.-B. Shu, Q.~Luo, Y.~Cheng, Z.-K. Hu, and M.-K. Zhou, ``A car-based portable atom gravimeter and its application in field gravity survey,'' {\em AIP Advances}, vol.~11, no.~11, 2021.

\bibitem{PetersDropping}
A.~Peters, K.~Y. Chung, and S.~Chu, ``Measurement of gravitational acceleration by dropping atoms,'' {\em Nature}, vol.~400, no.~6747, pp.~849--852, 1999.

\bibitem{inertialsensor}
M.~Wright, L.~Anastassiou, C.~Mishra, J.~Davies, A.~Phillips, S.~Maskell, and J.~Ralph, ``Cold atom inertial sensors for navigation applications,'' {\em Frontiers in Physics}, vol.~10, p.~994459, 2022.

\bibitem{gyro}
L.~Zhang, W.~Gao, Q.~Li, R.-B. Li, Z.~Yao, and S.~Lu, ``A novel monitoring navigation method for cold atom interference gyroscope,'' {\em Sensors}, vol.~19, p.~222, 01 2019.

\bibitem{gyro2}
A.~Hansen, Y.-J. Chen, J.~Kitching, and E.~Donley, ``Point-source atom interferometer gyroscope,'' Proceedings of the International School of Physics "Enrico Fermi", Varenna, IT, 2021-12-01 05:12:00 2021.

\bibitem{microwave}
F.~Zhou, F.~Jia, X.~Liu, Y.~Yu, J.~Mei, J.~Zhang, F.~Xie, and Z.~Zhong, ``Improving the spectral resolution and measurement range of quantum microwave electrometry by cold rydberg atoms,'' {\em Journal of Physics B: Atomic, Molecular and Optical Physics}, vol.~56, p.~025501, jan 2023.

\bibitem{computing}
M.~Weitz, ``Towards controlling larger quantum systems: from laser cooling to quantum computing,'' {\em IEEE Journal of Quantum Electronics}, vol.~36, no.~12, pp.~1346--1357, 2000.

\bibitem{PRLgravityclassical}
N.~Poli, F.-Y. Wang, M.~G. Tarallo, A.~Alberti, M.~Prevedelli, and G.~M. Tino, ``Precision measurement of gravity with cold atoms in an optical lattice and comparison with a classical gravimeter,'' {\em Phys. Rev. Lett.}, vol.~106, p.~038501, Jan 2011.

\bibitem{comparison}
Y.~Bidel, N.~Zahzam, A.~Bresson, C.~Blanchard, A.~Bonnin, J.~Bernard, M.~Cadoret, T.~E. Jensen, R.~Forsberg, C.~Salaun, S.~Lucas, M.~F. Lequentrec-Lalancette, D.~Rouxel, G.~Gabalda, L.~Seoane, D.~T. Vu, S.~Bruinsma, and S.~Bonvalot, ``Airborne absolute gravimetry with a quantum sensor, comparison with classical technologies,'' {\em Journal of Geophysical Research: Solid Earth}, vol.~128, no.~4, p.~e2022JB025921, 2023.

\bibitem{KasevichSRT}
M.~Kasevich and S.~Chu, ``Atomic interferometry using stimulated raman transitions,'' {\em Phys. Rev. Lett.}, vol.~67, pp.~181--184, Jul 1991.

\bibitem{kasevich1992measurement}
M.~Kasevich and S.~Chu, ``Measurement of the gravitational acceleration of an atom with a light-pulse atom interferometer,'' {\em Applied Physics B}, vol.~54, pp.~321--332, 1992.

\bibitem{Peters}
A.~{Peters}, K.~Y. {Chung}, and S.~{Chu}, ``{High-precision gravity measurements using atom interferometry},'' {\em Metrologia}, vol.~38, pp.~25--61, Feb. 2001.

\bibitem{rosi}
G.~Rosi, ``Precision gravity measurements with atom interferometry,'' {\em PhD University of Pisa}, 2012.

\bibitem{Zhou}
L.~Zhou, Z.-Y. Xiong, W.~Yang, B.~Tang, W.-C. Peng, Y.-B. Wang, P.~Xu, J.~Wang, and M.-S. Zhan, ``Measurement of local gravity via a cold atom interferometer,'' {\em Chinese Physics Letters}, vol.~28, no.~1, p.~013701, 2011.

\bibitem{g_measurement}
S.~Lellouch, K.~Bongs, and M.~Holynski, ``Using atom interferometry to measure gravity,'' {\em Contemporary Physics}, vol.~63, no.~2, pp.~138--155, 2022.

\bibitem{Gravimeter_dynamic_measurement}
C.-F. Huang, A.~Li, F.-J. Qin, J.~Fang, and X.~Chen, ``An atomic gravimeter dynamic measurement method based on kalman filter,'' {\em Measurement Science and Technology}, vol.~34, p.~015013, oct 2022.

\bibitem{gravimetryreview}
M.~de~Angelis, A.~Bertoldi, L.~Cacciapuoti, A.~Giorgini, G.~Lamporesi, M.~Prevedelli, G.~Saccorotti, F.~Sorrentino, and G.~M. Tino, ``Precision gravimetry with atomic sensors,'' {\em Measurement Science and Technology}, vol.~20, p.~022001, dec 2008.

\bibitem{Fountainmicrowave}
A.~E. Afanasiev, P.~I. Skakunenko, and V.~I. Balykin, ``Cold atom gravimeter based on an atomic fountain and a microwave transition,'' {\em JETP Letters}, vol.~119, pp.~84--88, Jan 2024.

\bibitem{rotation}
M.~Gilowski, C.~Schubert, T.~Wendrich, P.~Berg, G.~Tackmann, W.~Ertmer, and E.~M. Rasel, ``High resolution rotation sensor based on cold rubidium atoms,'' in {\em 2009 IEEE International Frequency Control Symposium Joint with the 22nd European Frequency and Time forum}, pp.~1173--1175, 2009.

\bibitem{wu2019gravity}
X.~Wu, Z.~Pagel, B.~S. Malek, T.~H. Nguyen, F.~Zi, D.~S. Scheirer, and H.~M{\"u}ller, ``Gravity surveys using a mobile atom interferometer,'' {\em Science advances}, vol.~5, no.~9, p.~eaax0800, 2019.

\bibitem{wu2014investigation}
B.~Wu, Z.~Wang, B.~Cheng, Q.~Wang, A.~Xu, and Q.~Lin, ``The investigation of a $\mu$gal-level cold atom gravimeter for field applications,'' {\em Metrologia}, vol.~51, no.~5, p.~452, 2014.

\bibitem{niebauer2007gravimetric}
T.~Niebauer, ``Gravimetric methods-absolute gravimeter: instruments concepts and implementation,'' {\em Geodesy}, vol.~3, pp.~43--64, 2007.

\bibitem{yu2006development}
N.~Yu, J.~Kohel, J.~Kellogg, and L.~Maleki, ``Development of an atom-interferometer gravity gradiometer for gravity measurement from space,'' {\em Applied Physics B}, vol.~84, pp.~647--652, 2006.

\bibitem{tino2013precision}
G.~M. Tino, F.~Sorrentino, D.~Aguilera, B.~Battelier, A.~Bertoldi, Q.~Bodart, K.~Bongs, P.~Bouyer, C.~Braxmaier, L.~Cacciapuoti, {\em et~al.}, ``Precision gravity tests with atom interferometry in space,'' {\em Nuclear Physics B-Proceedings Supplements}, vol.~243, pp.~203--217, 2013.

\bibitem{peters2001high}
A.~Peters, K.~Y. Chung, and S.~Chu, ``High-precision gravity measurements using atom interferometry,'' {\em Metrologia}, vol.~38, no.~1, p.~25, 2001.

\bibitem{singh2021single}
S.~Singh, B.~Jain, S.~P. Ram, V.~B. Tiwari, and S.~R. Mishra, ``A single laser-operated magneto-optical trap for rb atomic fountain,'' {\em Pramana}, vol.~95, pp.~1--5, 2021.

\bibitem{theoreticalSRT}
K.~Moler, D.~S. Weiss, M.~Kasevich, and S.~Chu, ``Theoretical analysis of velocity-selective raman transitions,'' {\em Physical Review A}, vol.~45, no.~1, p.~342, 1992.

\bibitem{Tinsley2019}
J.~N. Tinsley, {\em Construction of a Rubidium Fountain Atomic Interferometer for Gravity Gradiometry}.
\newblock PhD thesis, University of Liverpool, 2019.

\bibitem{kasevich1992}
M.~{Kasevich} and S.~{Chu}, ``{Measurement of the gravitational acceleration of an atom with a light-pulse atom interferometer},'' {\em Applied Physics B: Lasers and Optics}, vol.~54, pp.~321--332, May 1992.

\bibitem{vibrationcorrection}
M.~Guo, J.~Bai, D.~Hu, Z.~Tang, J.~You, R.~Chen, and Y.~Wang, ``A vibration correction system for cold atom gravimeter,'' {\em Measurement Science and Technology}, vol.~35, p.~035011, dec 2023.

\bibitem{egm2008}
N.~K. Pavlis, S.~A. Holmes, S.~C. Kenyon, and J.~K. Factor, ``The development and evaluation of the earth gravitational model 2008,'' {\em Journal of Geophysical Research: Solid Earth}, vol.~117, no.~B4, 2012.

\end{thebibliography}

\end{document}